\documentclass[superscriptaddress,floatfix,eqsecnum,aps,nofootinbib]{revtex4}
\usepackage{graphicx}
\usepackage{dcolumn}
\usepackage{bm}
\usepackage{graphics}
\usepackage{amsmath}
\usepackage{amssymb}
\usepackage{amscd}
\usepackage{afterpage}
\usepackage{float,times}
\usepackage{subfigure}
\usepackage{rotating}
\usepackage{multirow}
\usepackage{epsfig}
\usepackage{theorem}
\usepackage{moreverb}
\usepackage{euscript}
\usepackage{psfrag}

\begin{document}

\title{Soft supersymmetry breaking from stochastic superspace}

\author{Archil Kobakhidze}\email{archilk@unimelb.edu.au}
\affiliation{School of Physics, The University of Melbourne, Victoria 3010, Australia}
\author{Nadine Pesor}\email{npesor@student.unimelb.edu.au}
\affiliation{School of Physics, The University of Melbourne, Victoria 3010, Australia}
\author{Raymond R. Volkas}\email{raymondv@unimelb.edu.au}
\affiliation{School of Physics, The University of Melbourne, Victoria 3010, Australia}

\begin{abstract}
We propose a new realization of softly broken supersymmetric theories as theories defined on stochastic superspace. At the classical level, the supersymmetry breaking is parameterized in terms of a single (in general complex) mass parameter, $\xi$, describing the stochasticity of the Grassmannian superspace coordinates. In the context of the standard model with stochastic supersymmetry, the structure of the soft breaking terms has various characteristic features that can be tested in LHC experiments. Namely, at the classical level, the $B_{\mu}$ parameter, the universal soft trilinear coupling $A_0$, the universal gaugino mass $m_{1/2}$ and the universal scalar mass $m_0$ are given solely in terms of $\xi$; there are no other arbitrary parameters. The relations are $B_\mu = \xi^*$, $A_0 = 2\xi^*$, $m_{1/2} = |\xi|/2$ and $m_0 = 0$. At the quantum level, these relations hold at a certain scale $\Lambda$ which is a second free parameter.  The soft scalar masses, zero at tree-level, are induced radiatively through the renormalization group equations at one-loop. With this pattern of  soft breaking terms,  large supersymmetric contributions to FCNC processes are avoided. As a concrete illustration of the proposed formalism, we consider a minimal model, which is just the constrained MSSM with the stochastic superspace relations amongst the soft-breaking parameters imposed at the scale $\Lambda$.  We show that this theory is phenomenologically viable for a certain region in the $(\xi,\Lambda)$ parameter space. Some sensible extensions of the minimal model are then briefly discussed.
\end{abstract}

\maketitle

\section{Introduction}

Supersymmetry is the unique non-trivial extension of the relativistic Poincar\'e symmetry of spacetime. It leads to  field theories with improved ultraviolet behavior. Because of this, supersymmetry might have direct relevance to particle physics through its stabilizing of the electroweak scale under radiative corrections. If this is indeed the case, then the non-observation of supersymmetric particles means that  supersymmetry manifests at low energies in softly broken form: only those supersymmetry breaking operators are allowed which do not introduce new types of divergences (i.e.\ higher than logarithmic).   

Among renormalizable operators there are only three types of soft supersymmetry breaking terms: the mass term for the lowest scalar component of a chiral superfield, the mass term for the lowest fermionic component of a vector superfield and the scalar trilinear and bilinear couplings. In the supersymmetric version of the standard model, the generic supersymmetry breaking sector contains a large number of new unknown  parameters. Consequently, the  predictive power of the theory is compromised.  Furthermore, there are in general unacceptably large contributions to flavor changing neutral current processes due to the absence of a GIM-like mechanism in the sector of scalar partners (squarks) of ordinary quarks. To avoid this problem the squark soft breaking masses must be degenerate with high accuracy, unlike the observed hierarchical structure of quark-lepton masses.   To achieve universality in soft breaking masses is not an easy task. This typically requires rather complicated hidden and messenger sectors of supersymmetry breaking and, in many cases,  special flavor symmetries also need to be postulated.           

In this paper we suggest a conceptually  different view of softly broken supersymmetric theories.  Namely, we define a field theory on superspace where the Grassmannian coordinates are stochastic. With a suitably chosen probability distribution, to be defined below, the soft supersymmetry breaking parameters emerge upon averaging over the fluctuating superspace coordinates.  At the classical level, the supersymmetry breaking is parameterized through a single mass scale, $\xi$, that describes the stochasticity of the Grassmannian coordinates, and a very specific pattern of soft breaking terms emerges. Namely, at tree-level, the $B_{\mu}$ parameter, the universal soft trilinear coupling $A_0$, and the universal gaugino mass $m_{1/2}$ are given solely in terms of the stochasticity parameter $\xi$, with no other arbitrary parameters involved. At the quantum level, these relations are imposed at a certain scale $\Lambda$, which becomes a second free parameter.  The soft scalar masses, zero at tree-level, are induced radiatively through the renormalization group equations at one-loop. The resulting pattern of supersymmetry breaking is highly desirable phenomenologically since it is capable of resolving the above-mentioned supersymmetric flavor problem and it is extremely predictive.

Most phenomenological studies of supersymmetric extensions of the standard model are done using explicit soft supersymmetry breaking.  It is understood that one eventually wants to derive the soft breaking parameters from a fundamental theory of spontaneous or dynamical supersymmetry breaking, and to have that theory simultaneously explain the special pattern of soft breaking terms required for agreement with experimental constraints (degenerate squark masses, and so on).  The idea of stochastic superspace coordinates lies conceptually somewhere in-between a full fundamental theory and pure phenomenological parameterization.  It is much more specific than the latter, but it is not intended to be a fundamental theory either.  Our main point is that a certain line through the multi-dimensional parameter space of soft supersymmetry breaking turns out to be correlated with a non-trivial observation:  it corresponds to the hypothesis that superspace coordinates fluctuate and thereby break supersymmetry.  Furthermore, a section of this line produces phenomenologically realistic physics, and being so constrained it is ripe for experimental confirmation or falsification.  If the LHC were to discover a supersymmetric particle spectrum in accord with the stochastic superspace relations, that would motivate a new perspective on the desired fundamental theory.  That theory would have to produce fluctuating superspace coordinates at the effective low-energy level.  We shall not speculate here as to what such a theory could be like.  We begin at the beginning, namely phenomenology, and await the experimental results to come in the near future.

The rest of the paper is organized as follows. In the next section we define stochastic superspace and as a warm up exercise compute soft breaking parameters in the simple interacting Wess-Zumino model. In the section 3 we apply our formalism to the case of the minimal supersymmetric standard model (MSSM). We compute the low-energy soft breaking parameters using an approximate analytical solution to the renormalization group (RG) equations, and establish that this model is phenomenologically acceptable in a certain region of $(\xi,\Lambda)$ parameter space. We briefly mention possible extensions and then conclude.

\section{Stochastic superspace}

We consider N=1 8-dimensional superspace, with the Grassmannian coordinates $\theta$ and $\bar \theta$ being stochastic variables. One starts with the ordinary N=1 supersymmetric Lagrangian density expressed in terms of superfields.  The Lagrangian density in ordinary spacetime is then defined as an average of the superspace Lagrangian density over the different realizations of the  stochastic Grassmannian coordinates described by some probability distribution, ${\cal P}(\theta, \bar \theta)$. This probability distribution can be expanded as\footnote{We use conventions adopted in Ref. \cite{Lykken:1996xt}.}:
\begin{eqnarray}
{\cal P}(\theta, \bar \theta)=A+\theta^{\alpha} \Psi_{\alpha}+\bar \theta_{\dot{\alpha}} \bar \Xi^{\dot{\alpha}}
+\theta^{\alpha}\theta_{\alpha}B+\bar\theta_{\dot{\alpha}}\bar \theta^{\dot{\alpha}}C \nonumber \\
+\theta^{\alpha}\sigma^{\mu}_{~\alpha \dot{\beta}}\bar \theta^{\dot{\beta}}V_{\mu}+\theta^{\alpha}\theta_{\alpha}\bar \theta_{\dot{\alpha}}\bar \Lambda^{\dot{\alpha}}+
\bar \theta_{\dot{\alpha}}\bar \theta^{\dot{\alpha}} \theta^{\alpha } \Sigma_{\alpha}+\theta^{\alpha}\theta_{\alpha}\bar \theta^{\dot{\alpha}}\bar \theta_{\dot{\alpha}}D~,
\label{2}
\end{eqnarray}
where, $A$, $B$, $C$, $D$ and $V_{\mu}$ are c-numbers, while $\Psi$, $\bar \Xi$, $\bar \Lambda$ and $\Sigma$ are Grassmann numbers. We demand that the probability  measure (\ref{2}) satisfy the following conditions:

\begin{enumerate}
\item The normalization condition: 
\begin{equation}
\int d^2\theta d^2\bar \theta {\cal P}(\theta, \bar \theta)  = 1,~ \Longrightarrow ~  D=1~.
\label{3}
\end{equation}
\item 
That all Lorentz non-scalar moments vanish to ensure Lorentz invariance: 
\begin{equation}
\langle \theta \rangle= \langle \bar \theta \rangle=\langle\theta \bar \theta \rangle=\langle \theta^2 \bar \theta \rangle=\langle \theta\bar \theta^2  \rangle = 0,~ \Longrightarrow~   \Psi=\bar \Xi=V_{\mu}=\bar \Lambda=\Sigma = 0~.
\label{4}
\end{equation}
The non-vanishing moments are:  
\begin{eqnarray}
\langle \theta^{\alpha}\theta^{\beta}\rangle=\frac{1}{2\xi}\epsilon^{\alpha \beta},~ 
\langle \bar \theta_{\dot{\alpha}}\bar \theta_{\dot{\beta}}\rangle=\frac{1}{2\xi^{*}}\epsilon_{\dot{\alpha} \dot{\beta}},~ \langle \theta^{\alpha}\theta^{\beta}\bar \theta_{\dot{\alpha}}\bar \theta_{\dot{\beta}} \rangle=
\langle \theta^{\alpha}\theta^{\beta}\rangle\langle \bar \theta_{\dot{\alpha}}\bar \theta_{\dot{\beta}}\rangle=\frac{1}{4|\xi|^2}\epsilon^{\alpha \beta}\epsilon_{\dot{\alpha} \dot{\beta}} \nonumber \\
\Longrightarrow B=C^{*}=\frac{1}{\xi},~A=\frac{1}{|\xi|^2}~, 
\label{5}
\end{eqnarray}
where $\xi$ is a complex parameter of mass dimension.\footnote{In theories with exact U(1)$_R$ symmetry the phase of $\xi$ has no physical meaning and can be rotated away. Below in numerical estimations we simply take $\xi$ to be real.} 
\end{enumerate} 
Under these conditions we have the Hermitian probability measure, 
\begin{equation}
{\cal P}(\theta, \bar \theta )|\xi|^2\equiv \tilde {\cal P}(\theta, \bar \theta) =1+\xi^{*}(\theta \theta)+\xi
(\bar \theta \bar \theta) + |\xi|^2 (\theta \theta)(\bar \theta \bar \theta)~, 
\label{6}
\end{equation}  
for the stochastic Grassmann variables. We shall use the dimensionless  probability distribution $\tilde {\cal P}$ in what follows. The probability distribution in (\ref{6}) can be considered as the analog of the more familiar white noise distribution for bosonic stochastic variables. It is convenient to introduce $P(\theta)$ chiral  and $\bar P(\bar \theta)=P^{+}$  anti-chiral distributions as well, 
\begin{eqnarray}
P=1+\xi^* \theta^2, ~ \bar P=1+\xi \bar \theta^2~, \nonumber \\
\tilde {\cal P}=P\bar P,~~\int d^2\theta \frac{1}{\xi^*} P=\int d^2\bar \theta \frac{1}{\xi}\bar P=1~.
\end{eqnarray}  
The expression (\ref{6}) for the probability distribution represents a spurion superfield with non-zero F and D terms:
\begin{equation}
[ \tilde {\cal P} ]_{\rm F}= \xi^{*}\neq 0,~ [ \tilde {\cal P} ]_{\rm D}=|\xi|^2\neq 0~. 
\label{7}
\end{equation}
Therefore, upon averaging over the stochastic superspace, we expect to obtain a theory with softly broken F-type and D-type terms included. The parameter $\xi$ then defines the supersymmetry breaking scale.  

To see explicitly how this goes consider the simplest theory with a single chiral superfield $\Phi$, whose expansion in component fields is 
\begin{equation}
\Phi (x, \theta, \bar \theta)= \phi(x)+\sqrt{2}\theta \psi(x)+ \theta^2 F(x)+i \theta\sigma^{\mu}\bar \theta\partial_{\mu}\phi(x) +\frac{i}{\sqrt{2}}\theta^2 \partial_{\mu}\psi(x) \sigma^{\mu}\bar \theta- 
\frac{1}{4}\theta^{2}\bar \theta^2 \Box \phi(x)~.  
\label{8}
\end{equation} 
The Lagrangian of the model is the usual one modified by averaging over the distribution of the stochastic Grassmann coordinates,
\begin{equation}
L=\langle {\cal L}\rangle~,
\label{9} 
\end{equation}      
where ${\cal L}$ is the usual super-Lagrangian density which consists of two terms:  the kinetic term, which is the D-density, 
\begin{equation}
{\cal L}_{\rm kin}=\Phi^+\Phi~,
\label{10}  
\end{equation}
and the superpotential term, 
\begin{equation}
W= \frac{m}{2}\Phi^2+\frac{h}{3}\Phi^3
\label{11}
\end{equation}
(plus the anti-chiral superpotential $\bar W=W^{+}$) which is the F-density. 

The averaging of the kinetic Lagrangian (\ref{10}) results in
\begin{equation}
\langle {\cal L}_{\rm kin} \rangle =\int d^2\theta d^2\bar \theta P\bar P\Phi^{+}\Phi = L_{\rm kin-SUSY}+|\xi|^2\phi^*(x)\phi(x)+\xi^* \phi(x)F^{*}(x)+\xi \phi^{*}(x)F(x)~,
\label{12}
\end{equation}
where $L_{\rm kin-SUSY}=\int d^2\theta d^2\bar \theta \Phi^{+}\Phi $ is the usual supersymmetric kinetic Lagrangian.  Solving for the auxiliary field (ignoring the superpotential for the moment), 
\begin{equation}
F(x)=-\xi^* \phi~,
\label{13}
\end{equation}
and plugging (\ref{13}) back into (\ref{12}) we obtain that all terms proportional to $\xi$ cancel with each other, and the remaining Lagrangian is invariant under the on-shell N=1 supersymmetry transformations. 

Let us now compute the potential density:
\begin{equation}
V=\langle W \rangle +{\rm h.c.}=V_{\rm SUSY}+\left(\frac{\xi^* m}{2}\phi^2+\frac{\xi^* 
 h}{3}\phi^3 +{\rm h.c.} \right)~,
\label{15}
\end{equation}
where $V_{\rm SUSY}$ is invariant under N=1 supersymmetry transformations. The F-term now reads: 
\begin{equation}
F=-\xi^* \phi+m^{*}\phi^{*}+h^{*}\phi^{*2}~.
\label{16}
\end{equation}
Substituting (\ref{16}) back into the total Lagrangian ($\langle L_{\rm kin}+W\rangle$), we obtain
\begin{equation}
L=L_{\rm on-shell - SUSY}-\left(\frac{\xi^* m}{2}\phi^2+\frac{2\xi^* h}{3}\phi^3 +{\rm h.c.} \right)~,
\label{17}
\end{equation}
where $L_{\rm on-shell - SUSY}$ is the familiar on-shell supersymmetric Lagrangian of the interacting Wess-Zumino model. The remaining terms in (\ref{17}) break supersymmetry softly. Thus we have  a Majorana fermion $\psi$ with mass $|m|$ and two scalar fields with masses $\sqrt{|m|^2+|\xi^* m|}$ and  $\sqrt{|m|^2-|\xi^* m|}$. To avoid tachyonic states we must assume $|\xi | \leq |m|$. The supertrace STrM$^2$   is zero.  

The above simple case captures some key aspects of more generic theories with stochastic supersymmetry. Namely, the averaged D-density, 
\begin{equation}
L_{\rm kin-gauge}=\langle {\rm Tr} \Phi^{+}e^{2gV}\Phi \rangle~,
\label{18}
\end{equation}
of a charged scalar superfield $\Phi$ interacting with a gauge superfield (in the Wess-Zumino gauge),
\begin{equation}
V(x, \theta, \bar \theta)=-\theta \sigma^{\mu}\bar \theta A_{\mu}(x)+i\theta^{\alpha}\theta_{\alpha}\bar\theta_{\dot \alpha}\bar \lambda^{\dot \alpha}(x)-i\bar \theta_{\dot \alpha}\bar \theta^{\dot \alpha}\theta^{\alpha}\lambda_{\alpha}(x)+\frac{1}{2}\theta^2\bar \theta^2 D(x)~,
\label{19}
\end{equation}
does not produce supersymmetry breaking terms. Soft breaking terms emerge from the (gauge invariant) superpotential only.\footnote{Gauge invariance requires the introduction of more than one chiral superfield.} 

\section{Standard Model in stochastic superspace}

The structure of the soft breaking terms in the context of the minimal standard model with stochastic supersymmetry should now be clear.  They all come from the F-densities. For example, averaging of the superpotential (with R-parity conservation being assumed),
\begin{equation}
W_{\rm SM}=\mu H_u H_d+\hat y^{\rm up}QU^cH_u+\hat y^{\rm down}QD^cH_d+\hat y^{\rm lept}LE^cH_d
\label{20}
\end{equation}   
results in the soft breaking terms
\begin{equation}
L_{\rm soft-scalar}=-\xi^*\mu \tilde H_u\tilde H_d-2\xi^*\left[    
\hat y^{\rm up}\tilde Q\tilde U^c\tilde H_u+\hat y^{\rm down}\tilde Q\tilde D^c\tilde H_d+\hat y^{\rm lept}\tilde L\tilde E^c\tilde H_d
\right] + {\rm h.c.}
\label{21}
\end{equation}
where fields with tildes denote the scalar components of the corresponding chiral superfields (in a self-evident notation). Soft breaking masses for the gauginos $\lambda^{(i)}(x)$ appear upon averaging of the gauge-kinetic F-densities, 
\begin{equation}
L_{\rm gauge}=\langle \frac{1}{2}\sum_i {\rm Tr}W^{(i)\alpha}W^{(i)}_{\alpha}\rangle +{\rm h.c.}=
\left[ \frac{1}{2}\sum_i {\rm Tr} W^{(i)\alpha}W^{(i)}_{\alpha}\right]_F-\frac{\xi^*}{2}\sum_i{\rm Tr}\lambda^{(i)}\lambda^{(i)}+{\rm h.c.}
\label{22}
\end{equation}
where
\begin{equation}
W^{(i)\alpha}=-i\lambda^{(i)\alpha}(y)+\theta^{\alpha}D^{(i)}(y)+\theta^{\alpha}\sigma^{\mu\nu~\beta }_{\alpha}F_{\mu\nu}^{(i)}(y)-\theta^2\bar \sigma^{\mu~\dot \beta\alpha}D^{(i)}_{\mu}\bar \lambda^{(i)}_{\dot \beta}(y)~,
\label{23}
\end{equation}
are the field-strength spinor superfields for SU(3) ($i=3$), SU(2) ($i=2$) and U(1)$_{Y}$ ($i=1$) subgroups of the standard model gauge group, $D^{(i)}_{\mu}$ are the corresponding covariant derivatives and $y^{\mu}=x^{\mu}+i\theta\sigma^{\mu} \bar \theta$ is the chiral coordinate. The fundamental-representation generators are normalized as ${\rm Tr }T^aT^b=\frac{1}{2}\delta^{ab}$. 

Summarizing, we have obtained the following tree-level soft breaking terms:
\begin{itemize}
\item The bilinear Higgs soft term with $B_{\mu}=\xi^{*}$.
\item The trilinear scalar soft terms proportional to the Yukawa couplings, with the universal constant $A_0=2\xi^{*}$.
\item The universal gaugino masses, $m_{1/2}=\frac{1}{2}|\xi|$.
\item The scalar soft masses are absent, $m_0^2=0$.
\end{itemize}
At the quantum level, the above soft breaking terms are, of course, renormalized, so the relations hold only at some energy scale $\Lambda$ which joins $\xi$ as a free parameter in the theory.

Remarkably, the pattern of the soft breaking terms represents a particular case of the more general pattern of the so-called constrained minimal supersymmetric Standard Model (CMSSM). The CMSSM soft breaking terms  are  usually  motivated by spontaneous supersymmetry breaking within {\it minimal}  N=1 supergravity.\footnote{In the minimal version of N=1 supergravity one assumes the minimal flavor-blind structure for the K\"{a}hler potential, which does not follow from any symmetry principle. Hence, the phenomenologically desirable pattern of the soft breaking terms in the CMSSM is arguably not so well-motivated theoretically.}   The model is defined through  the following set of extra parameters (beyond those of the standard model): $(m_0,~A_{0},~m_{1/2},~\tan\beta,~{\rm sgn}{\mu})$.\footnote{It is customary to express the modulus of $\mu$ through the known standard model parameters and the bilinear soft parameter $B_{\mu}$ through  $\tan\beta \equiv \langle H_U^0\rangle/\langle H_D^0\rangle$, using the extremum conditions for the Higgs potential. } This model has been extensively studied in the literature and the constraints on the soft breaking parameters have been obtained from various astrophysical and collider experimental data.

Obviously, our model is even more constrained than the CMSSM.  Nevertheless, we shall show that it is phenomenologically viable in an appropriate region of $(\xi,\Lambda)$ parameter space.  To do this, we solve the renormalization group (RG) equations and compute the sparticle spectrum at the weak scale as a function of $\xi$, $\Lambda$ and ${\rm sgn}{\mu}$.

The RG equations are, of course, a complicated set of coupled differential equations which in general require numerical calculations. However, in certain regimes an analytical solution to an approximation of the one-loop RG equations is available \cite{Kazakov:2000us}. We consider the regime of small/moderate $\tan\beta$  where the only important Yukawa coupling is the top-Yukawa coupling $y_t$ (the other Yukawa couplings being set to zero in the RG equations). We focus on the regime $M_{\rm GUT}  < \Lambda < M_P$, where $M_{\rm GUT} \simeq 2\cdot 10^{16} {\rm GeV}$ is the putative grand-unification scale and $M_P \simeq 10^{19}\ {\rm GeV}$ is the Planck mass.  In this regime we have obtained an approximate solution for $\tan\beta$,\footnote{The grand unified coupling is taken as $\alpha_{\rm GUT}\approx 1/24.3$. Other input parameters are taken as follows: the running top mass $m_{t}\approx 160.4$ GeV (corresponding to the pole mass $m_t^{\rm pole}\approx 170.9$ GeV \cite{:2007bxa}), $\sin^2\theta_W(M_Z)\approx 0.2315$, and we also assume that $\mu$ is positive. All running parameters are evaluated at the Z-pole, $M_Z\approx 91.2$ GeV.} 
\begin{equation}
\tan\beta \approx 5.9\ ({\rm for}\ \Lambda = M_{\rm GUT})\ {\rm and}\ 7.5\ ({\rm for}\ \Lambda = M_P),
\label{25}
\end{equation}
where we have taken $\xi < 0$. These values of $\tan\beta$ turn out to be largely insensitive to the parameter $\xi$. For positive $\xi$ the values of $\tan\beta$ are low, $\tan\beta \approx 2.8$ for $\Lambda = M_{\rm GUT}$ and $2.4$ for $\Lambda = M_P$. In this case the predicted mass for the lightest CP-even Higgs mass fails to satisfy the LEP limit \cite{Barate:2003sz}. 
The one-loop soft scalar masses for sleptons and the first two generations of squarks (which with good accuracy can be considered as the mass eigenstates) are:
\begin{eqnarray}
m^2_{\tilde u_L}\approx (1.70, 2.09)\xi^2+(4M_W^2-M_Z^2)\cos(2\beta)/6, \nonumber \\
m^2_{\tilde d_L}\approx (1.70, 2.09)\xi^2-(2M_W^2+M_Z^2)\cos(2\beta)/6, \nonumber \\
m^2_{\tilde u_R}\approx (1.60, 1.97)\xi^2-2(M_W^2-M_Z^2)\cos(2\beta)/3, \nonumber \\
m^2_{\tilde d_R}\approx (1.59, 1.96)\xi^2+(M_W^2-M_Z^2)\cos(2\beta)/3, \nonumber \\
m^2_{\tilde \nu_L}\approx (0.12, 0.15)\xi^2+M_Z^2\cos(2\beta)/2,        \nonumber \\
m^2_{\tilde e_L}\approx (0.12, 0.15)\xi^2-(2M_W^2-M_Z^2)\cos(2\beta)/2, \nonumber \\
m^2_{\tilde e_R}\approx (0.038, 0.041)\xi^2+(M_W^2-M_Z^2)\cos(2\beta)~, 
\label{24}
\end{eqnarray}
where the first number in parentheses is for $\Lambda = M_{\rm GUT}$ and the second for $\Lambda = M_P$.
The third generation soft masses are different due to the contribution from the large top-Yukawa coupling, 
\begin{eqnarray}
m^2_{\tilde t_L}\approx m^2_{\tilde u_L}+m_t^2+\frac{\Delta}{6}, \nonumber \\
m^2_{\tilde b_L}\approx m^2_{\tilde d_L}+\frac{\Delta}{6},  \nonumber \\
m^2_{\tilde t_R}\approx m^2_{\tilde u_R}+m_t^2+\frac{\Delta}{3}, \nonumber \\
m^2_{\tilde b_R}\approx m^2_{\tilde d_R}, 
\label{thirdgen}
\end{eqnarray}
with sizable mixing in the stop sector. The explicit expression for $\Delta$ can be found in \cite{Kazakov:2000us}. For our choice of soft parameters at the unification scale, $\Delta$ evaluated down at the $M_Z$-scale is: $\Delta(M_Z)\approx (-3.36,-3.92)\xi^2$. Other parameters are found to be: $A_{t}(M_Z)\approx (1.82,1.96)\xi$, $B_{\mu}(M_Z)\approx (0.23,0.19)\xi$ (recall, $\xi<0$) and $\mu^2(M_Z)\approx (1.61,1.85)\xi^2-M_Z^2/2$. The physical stop masses are:
\begin{equation}
m_{\tilde t_{1,2}}^2\approx \frac{1}{2}\left[ (m_{\tilde t_L}^2+m_{\tilde t_R}^2)\mp  \sqrt{(m_{\tilde t_L}^2-m_{\tilde t_R}^2)^2+4m_t^2(A_t-\mu \cot\beta)^2} \right]~.
\label{26}
\end{equation} 

\begin{figure}
\centering
\includegraphics[angle=0, width=0.75\textwidth]{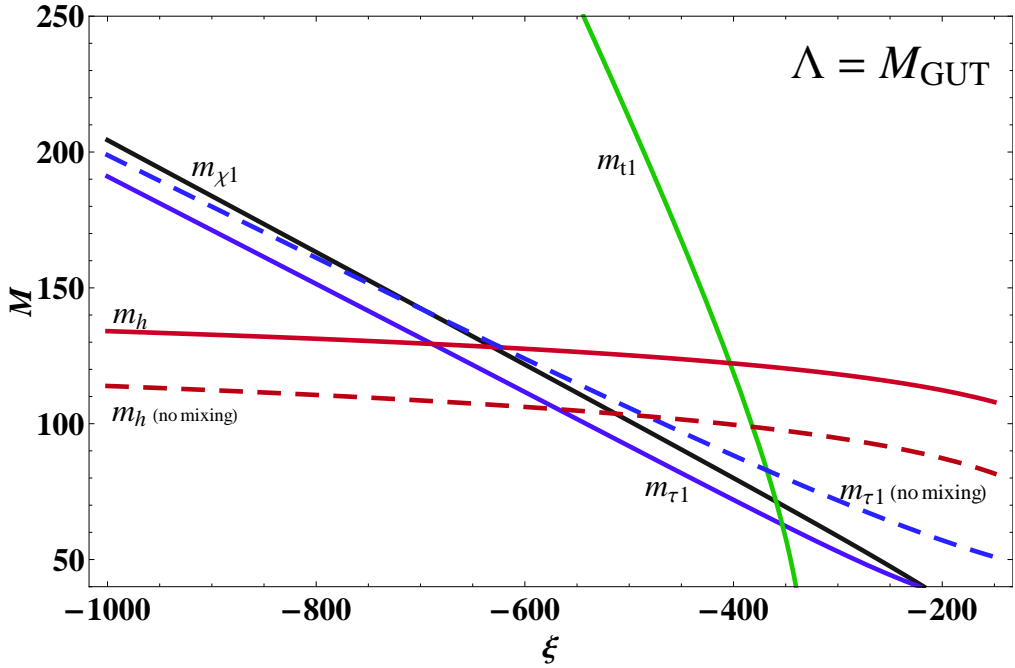}
  \caption{\small For the choice $\Lambda = M_{\rm GUT}$, this plot shows the masses (in GeV) of the lightest stau (marked as `$m_{\tau 1}$'), the lightest neutralino (`$m_{\chi 1}$') the 
lightest CP-even Higgs boson (solid line marked as `$m_h$') and the lightest stop (marked as `$m_{t1}$') as functions of $\xi$ (also in GeV).  To signify the importance of the mixing in the stop and stau sectors we have plotted also the $\xi$ dependence of $m_h$ and $m_{\tau 1}$ for the case when the corresponding mixings are ignored [dashed lines marked as `$m_h$ (no mixing)' and `$m_{\tau 1}$ (no mixing)'].  Because the lightest superpartner here is the stau, the $\Lambda = M_{\rm GUT}$ possibility is ruled out for the minimal model. }
  \label{figure1}
\end{figure}

Finally, we estimate the mass of the lightest CP-even Higgs boson in the decoupling limit, $m_A \gg M_Z$, of the one-loop leading logarithmic approximation \cite{Haber:1996fp}, 
\begin{eqnarray}
m_h^2\approx M_Z^2\cos^2(2\beta) +\frac{3}{4\pi^2}\frac{m_t^4}{v^2}
\log\left( \frac{M_L M_R}{m_t^2} \right) \nonumber \\ 
+\frac{3}{4\pi^2}\frac{m_t^4}{v^2}\left[(A_t-\mu\cot\beta)^2\left(h(M_L^2,M_R^2)+\frac{(A_t-\mu\cot\beta)^2}{2}g(M_L^2, M_R^2)\right)  \right],
\label{29}
\end{eqnarray}
where $v \simeq 174$ GeV, $h(a,b)\equiv \frac{1}{a-b}\ln\left(\frac{a}{b}\right)$, $g(a,b)\equiv \frac{1}{(a-b)^2}\left[
2-\frac{a+b}{a-b}\ln\left(\frac{a}{b}\right) \right]$, and $M_L^2=m^2_{\tilde t_L}-m_t^2-(4M_W^2-M_Z^2)\cos(2\beta)/6$, $M_R^2=m^2_{\tilde t_R}-m_t^2+2(M_W^2-M_Z^2)\cos(2\beta)/3$.  Remarkably enough, we are very close to the so-called maximal mixing scenario (see, e.g., \cite{Carena:2005ek}), where the lightest Higgs mass is enhanced significantly due to the mixing in the stop sector. This can be seen from Figure~\ref{figure1}, where we have plotted $m_h$ as a function of $\xi$ with mixing (solid line marked as `$m_h$') and without mixing (dashed line marked as `$m_h$ (no mixing)'). When the mixing is neglected $m_h$ fails to satisfy the LEP bound. 

However, a similar large mixing is present in the stau sector as well, due to the universality of the trilinear soft breaking parameter $A_0$ at the scale $\Lambda$. For the case $\Lambda = M_{\rm GUT}$ this results in the lightest stau mass eigenstate being lighter than the lightest neutralino for $\xi \lesssim -190$ GeV, as can be seen from Figure~\ref{figure1}. Note also that the region $\xi \gtrsim -380$ GeV is excluded by the upper bound on the lightest stop $m_{\tilde t_1}>95.7$ GeV (at $95\%$ C.L.) \cite{Amsler:2008zz}. Thus, in the absence of R-parity violation, the lightest stau is a stable particle, and this is excluded by observations. 

In the above calculation we have assumed $\mu > 0$. For negative $\mu$, we obtain a somewhat heavier stau. However, it is still the LSP for the region of  $\xi$'s where the bound on the mass of the lightest stop is satisfied.  Finally, as has been mentioned earlier, for positive $\xi$'s (and for either sign of $\mu$) the lightest CP-even Higgs boson is too light to satisfy the LEP bound. We conclude that the minimal model with $\Lambda = M_{\rm GUT}$ is excluded experimentally.   

\begin{figure}
\centering
\includegraphics[angle=0, width=0.75\textwidth]{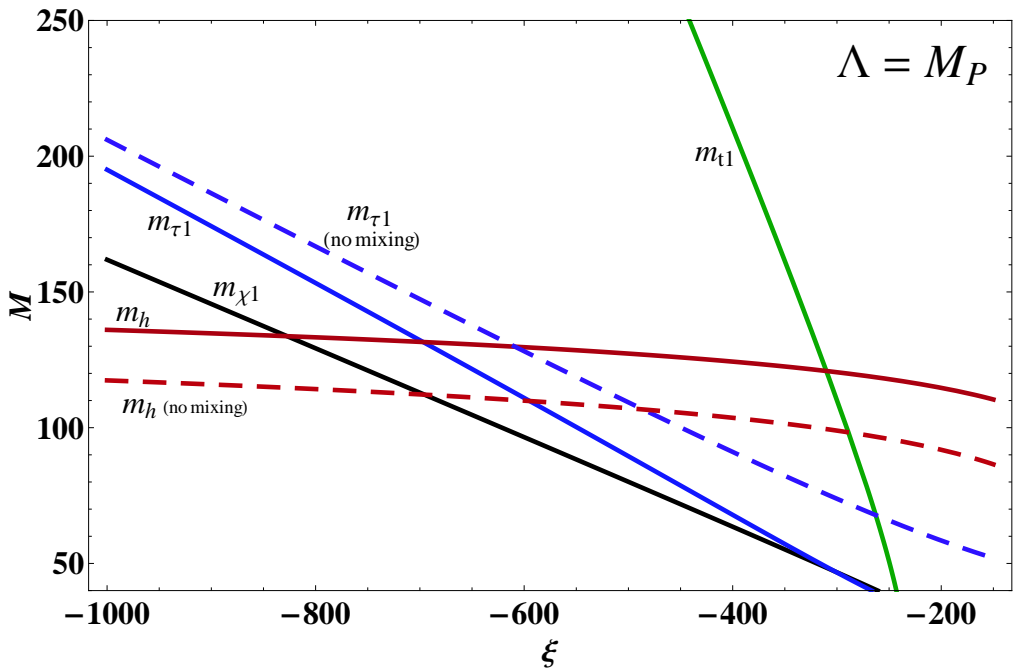}
  \caption{\small For the choice $\Lambda = M_P$, this plot shows the masses (in GeV) of the lightest stau (marked as `$m_{\tau 1}$'), the lightest neutralino (`$m_{\chi 1}$') the 
lightest CP-even Higgs boson (solid line marked as `$m_h$') and the lightest stop (marked as `$m_{t1}$') as functions of $\xi$ (also in GeV).  To signify the importance of the mixing in the stop and stau sectors we have plotted also the $\xi$ dependence of $m_h$ and $m_{\tau 1}$ for the case when the corresponding mixings are ignored [dashed lines marked as `$m_h$ (no mixing)' and `$m_{\tau 1}$ (no mixing)'].  The lightest superpartner here is the neutralino for $\xi \lesssim -300$ GeV. We have assumed $\mu > 0$. }
  \label{figure2}
\end{figure}

As we raise $\Lambda$ above $M_{\rm GUT}$, the stau becomes more massive relative to the lightest neutralino.  Let us repeat the above calculations for the other extreme case, $\Lambda = M_P$.  The results are displayed in Fig.~\ref{figure2}.  We see that for $\xi \lesssim -300$ GeV, the LSP is now the lightest neutralino, so the phenomenological problems of the $\Lambda = M_{\rm GUT}$ case are absent.

\begin{figure}
\centering
\includegraphics[angle=0, width=0.8\textwidth]{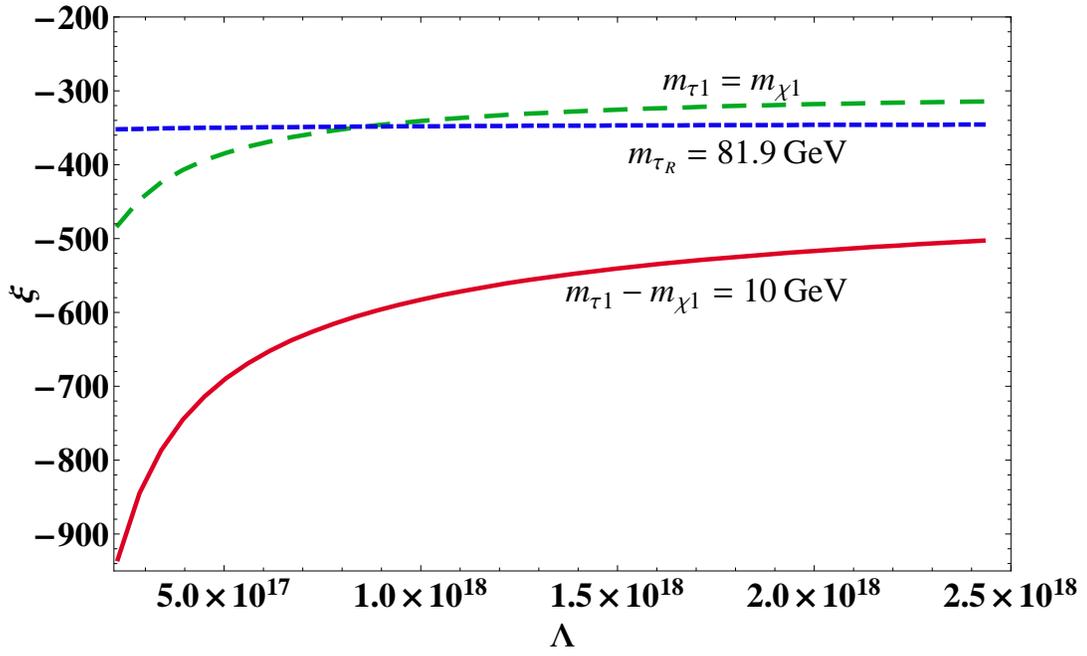}
  \caption{\small The broadly acceptable region of $(\xi,\Lambda)$ parameter space lies below the long- and short-dashed lines, with values between the long-dashed and solid lines favored on the basis of neutralino dark matter abundance calculations. This is the so-called stau co-annihilation regime.}
  \label{figure3}
\end{figure}

\begin{figure}
\centering
\includegraphics[angle=0, width=0.75\textwidth]{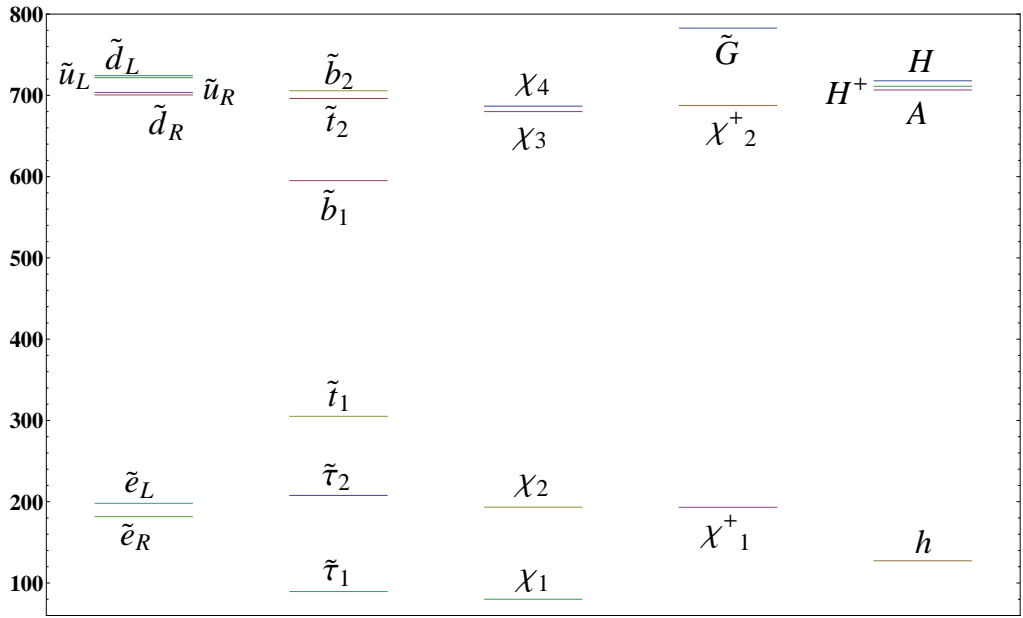}
  \caption{\small The complete superparticle spectrum for the phenomenologically-viable example $\xi = -500$ GeV and $\Lambda = M_P$.}
  \label{figure4}
\end{figure}

For a given value of $\xi$, a phenomenologically acceptable outcome is obtained for $\Lambda$ above a certain value that lies between $M_{\rm GUT}$ and $M_P$.  Figure \ref{figure3} illustrates this point.  The $(\xi,\Lambda)$ region above the long-dashed line is ruled out because the stau is the LSP.  The region above the short-dashed line is also ruled out because of the experimental lower bound on the right-handed stau mass.  The region below both the long- and short-dashed lines is broadly acceptable, with the region between the long-dashed and solid lines depicting the parameter space where the neutralino is up to $10$ GeV lighter than the lightest stau.  In this regime, co-annihilation processes in the early universe ensure the correct cosmological neutralino dark matter abundance is obtained.  The complete superparticle spectrum for the phenomenologically-viable example $\xi = -500$ GeV and $\Lambda = M_P$ is shown in Fig.~\ref{figure4}.

Finally, let us note that the minimal model has to be extended in order to incorporate nonzero neutrino masses. In such extensions, it is very plausible that the $\Lambda = M_{\rm GUT}$ choice can be phenomenologically viable. For example, one can replace the R-parity conservation of the minimal model by the less restrictive baryon B-parity conservation condition \cite{Ibanez:1991pr}. Then a new set of interactions are allowed which render the stau LSP unstable. The very same interactions generate neutrino masses radiatively. Other potentially viable models for lower values of $\Lambda$ are the next-to-minimal supersymmetric model and models which include sterile neutrino superfields. Finally, one can also consider the models where ordinary F- or/and D-term spontaneous supersymmetry breakings are incorporated within the stochastic supersymmetry formalism.

\section{Conclusion and outlook}

In this paper we have proposed a conceptually different approach to softly broken supersymmetric theories. We have demonstrated that a field theory defined on stochastic superspace is equivalent to a supersymmetric theory with a specific and constrained set of soft breaking terms. We have explicitly examined the case of the R-parity conserving MSSM where the Higgs bilinear soft parameter $B_{\mu}$, the universal trilinear scalar coupling $A_0$ and universal gaugino mass  are given solely in terms of stochasticity parameter $\xi$, and the universal scalar mass $m_0$ is zero.  At the quantum level, these relations hold at a scale $\Lambda$ which thus becomes a second parameter. As a result of these features we have been able to predict $\tan\beta$, as per Eq.~(\ref{25}), to express low energy soft masses through $\xi$, and to derive the entire superparticle spectrum.  Even though the theory is very constrained, it is nevertheless phenomenologically viable for $\Lambda > M_{\rm GUT}$ with $\xi$ negative and of a few $100$'s of GeV in magnitude.  The superparticle spectra predicted by stochastic superspace can be experimentally tested at the LHC.  Should those results turn out to be consistent with a stochastic superspace explanation, then a new perspective will have been gained on the as yet unknown fundamental theory of spontaneous or dynamical supersymmetry breaking.

\acknowledgments{ 
We would like to thank Csaba Balazs, Tony Gherghetta and Ilia Gogoladze for many valuable discussions and comments on the phenomenology of the minimal model presented here.  This work was supported by the Australian Research Council.}

\end{document}